\documentclass[nobm,figures,cite]{epl}

\title{Rocking bistable systems: use and abuse of Linear Response
Theory} \shorttitle{Rocking bistable systems: use and ...}
\author{J.~Casado-Pascual\inst{1,2} \and J.~G\'omez-Ord\'o\~nez\inst{1}
\and M.~Morillo\inst{1} \and P.~H\"anggi\inst{2}} \institute{ \inst{1}
F\'{\i}sica Te\'orica, Universidad de Sevilla - Apartado de Correos
1065, Sevilla 41080, Spain\\ \inst{2} Institut f\"ur Physik,
Universit\"at Augsburg - Universit\"atsstra\ss e 1, D-86135 Augsburg,
Germany } \shortauthor{J.~Casado-Pascual \etal}
 {Classical statistical mechanics}

\begin{document}

\maketitle

\begin{abstract}
The response of a nonlinear stochastic system driven by an external
sinusoidal time dependent force is studied by a variety of numerical and
analytical approximations. The validity of linear response theory is put
to a critical test by comparing its predictions with numerical
solutions over an extended parameter regime of driving
amplitudes and frequencies. The relevance of the driving frequency
for the applicability of linear response theory is explored.
\end{abstract}


The response of dissipative physical systems to small amplitude external
perturbations is usually described with the powerful tools of linear
response theory (LRT)\cite{kubo57}, as it is generally accepted that the
effect of the perturbation can be described in terms of small deviations
from the behavior of the unperturbed system.  In particular, for long times
and for systems which in the absence of driving reach an equilibrium
distribution, LRT provides an approximate expression for the probability
distribution obtained by keeping just the linear terms in a series
expansion in the external
amplitude. The purpose of the present letter is to point out the
relevance of parameters other than the amplitude of the driving
force, for the validity of LRT. We will show that
for a periodic external force, the validity of LRT depends not just on the
amplitude of the driving term but also crucially on its frequency.

Let us consider a system characterized by a single degree of freedom,
$x$, whose time evolution is governed by the nonlinear Langevin equation
(in dimensionless form),
\begin{equation}
\dot{x}(t)=x(t)-x^{3}(t)+A\cos \Omega t+\eta(t),
\end{equation}
where $A\cos\Omega t$ represents an external signal and
$\eta(t)$ is a Gaussian white noise with zero average and
$\langle \eta(t)\eta(s)\rangle = 2D\delta(t-s)$.  The corresponding linear
Fokker-Planck equation (FPE) for the probability density $P(x,t)$ reads
\begin{equation}
\label{lfpe}
\frac{\partial P}{\partial t}=\frac{\partial}{\partial
x}\big\{(-x+x^{3}- A\cos\Omega t)P\big\}+D\frac{\partial^{2}P}{\partial
x^{2}}.
\end{equation}                                    
The unperturbed system has an equilibrium distribution of the form
\begin{equation}
\label{peq}
P_{eq}(x)=N \exp \left (-\frac {U_0(x)}{D}\right ),
\end{equation}
where $N$ is a normalization constant and $U_0(x)$ is the unperturbed
potential 
\begin{equation}
U_0(x)=-\frac {x^2}2+\frac {x^4}4.
\end{equation}
This potential has two minima located at $x_m=\pm 1$ and a maximum at
$x_M=0$, with a barrier height of $0.25$. The potential $U_0(x)-Ax\cos
\Omega t$ loses its bistable character for $A \ge A_T=\sqrt{4/27}$.

The analysis of the dynamics is simplified by making use of two
important theorems: the H-theorem, which ensures the existence of a
unique long time distribution function $P_{\infty}(x,t)$
\cite{Lebowitz,Risken} and the Floquet theorem, which guarantees that
$P_{\infty}(x,t)$ is periodic in time with the same period as the
external force \cite{junhan91}. For the system at hand, the symmetry of
$U_0(x)$ implies the following properties for the long time unique
solution of the FPE: $P_\infty(-x,t;-A)=P_\infty(x,t;A)$ and
$P_\infty(-x,t;A)=P_\infty(x,t+T/2;A)$, where $T=2\pi/\Omega$ and we
have indicated explicitely the dependence of $P_{\infty}$ on $A$. Using
the Fourier expansion,
\begin{equation}
\label{fourier}
P_{\infty}(x,t;A)=\sum_{m=-\infty}^\infty H_m(x;A) e^{im\Omega t}
\end{equation}
the first property leads to $H_m(x;A)=H_m(-x;-A)$, while the second one
implies that $H_m(x;A)=(-1)^m H_m(-x;A)$. From both of them, we obtain
$H_m(x;-A)=(-1)^m H_m(x;A)$. It then follows immediately that the odd
moments of the distribution, $\langle x^n(t)
\rangle_\infty,\;n=1,3,\ldots$ can be written as Fourier series
containing only odd harmonics as the even harmonics vanish due to the
symmetries above. Analogously, even moments $\langle
x^p(t)\rangle_\infty,\;p=0,2,\ldots$ contain just even harmonics in
their Fourier series expansions\cite{Haenggi93}. Inserting the Fourier
expansion, eq.~(\ref{fourier}), into the FPE, an infinite set of
equations for the coefficients $H_m(x;A)$ is obtained. Inspection of the
set indicates that if $H_m(x,A)$ is expanded in powers of $A$, it can
not contain powers smaller than $A^{|m|}$.  From the above general
considerations, we have, in particular, for the first two moments,
\begin{eqnarray}
\langle x(t)\rangle_\infty &=& \sum_{n\,odd} M_n(A) e^{in\Omega t}
=2\sum_{n>0,odd}^\infty |M_n(A)| \cos (n \Omega t- \phi_n)\nonumber \\
&=&2\sum_{n>0,odd}^\infty
|M_n(A)| \left ( \cos \phi_n \cos n \Omega t + \sin  \phi_n \sin n \Omega
t \right )
\end{eqnarray}
with $M_n(A)=c_n^{(0)}A^{|n|}+c_n^{(2)}A^{|n+2|}+\ldots$, and
\begin{equation}
\langle x(t)^2\rangle_\infty = \sum_{p\;even} L_p(A) e^{ip\Omega t}
\end{equation}  
with $L_p(A)=b_p^{(0)}A^{|p|}+b_p^{(2)}A^{|p+2|}+\ldots$.
 
The exact analytical expression for $P_{\infty}(x,t)$ is unknown. LRT
amounts to write
\begin{equation}
\label{appr}
P_{\infty}(x,t)=P_{eq}(x) + A P_1^{(1)}(x,t),
\end{equation}
with $P_1(x,t)$ obtained from a first order perturbation analysis of the
FPE; see \cite{junhan91} and the Appendix of \cite{gamhan98} for details
\footnote{Note that the plus signs in (A.23) of Ref. \cite{gamhan98}
should read minus. This in turn yields a minus sign on the
right hand side in (A.28).}

\begin{equation}
\label{p1}
A P_1^{(1)}(x,t)=-\sum_{n=1}^\infty \frac A{\lambda_n^2+\Omega^2} \left [
\lambda_n \cos \Omega t + \Omega \sin \Omega t \right ] d_n \varphi_n(x),
\end{equation}
where $\varphi_n(x)$ are the right eigenstates of the unperturbed FP
operator and $\lambda_n$ the corresponding eigenvalues. The coefficients
$d_n$ are $\langle \varphi_n|\partial /\,\partial
x|\varphi_0\rangle$.  It follows from eqs.~(\ref {appr}) and (\ref{p1})
that the average value $ \langle x(t) \rangle_\infty^{LRT}$ is given by
\begin{equation}
\label{avg}
\langle x(t)\rangle^{LRT}_\infty = a_1 \cos (\Omega t - \phi_1^{LRT}).
\end{equation}
The explicit calculation of the amplitude, $a_1$, and phase lag,
$\phi_1^{LRT}$, requires the knowledge of the spectrum of the unperturbed
system. For the bistable system at hand, no exact analytical expressions
for the eigenfunctions and eigenvalues exist, although useful
approximate expressions are known
\cite{hanggi82,dykman93}. Alternatively, the amplitude and phase lag
\cite{gamhan98,junhan93,gommor94}  can
be obtained from the response function. Using the two-mode
approximation of Jung and H\"anggi \cite{junhan93}, we write
\begin{equation}
\langle x(t)\rangle^{LRT}_\infty = b_1 \cos(\Omega t- \beta_1) +b_2
\cos(\Omega t- \beta_2)
\end{equation}
where the first term on the right hand side is due to the interwell
hops, while the second one describes the influence of intrawell
dynamics. It is convenient to cast the expression for $\langle
x(t)\rangle^{LRT}_\infty$ as in eq.~(\ref{avg}), and within the two-mode
 approximation,  we get for the amplitude  
\begin{equation}
\label{amp}
a_1= \frac AD \left [ \frac {g_1^2 \lambda_1^2}{\lambda_1^2+\Omega^2}
+\frac {g_2^2 \alpha^2}{\alpha^2+\Omega^2} +\frac{2g_1g_2 \lambda_1
\alpha (
\lambda_1\alpha+\Omega^2)}{(\lambda_1^2+\Omega^2)(\alpha^2+\Omega^2)}
\right ]^{\frac 12}
\end{equation}
while the phase lag of the response with respect to the input signal,
$0 \le \phi_1^{LRT} \le \pi/2$, is given by 
\begin{equation}
\label{fase}
\phi_1^{LRT}=\arctan \frac{\frac
{g_1\lambda_1\Omega}{\lambda_1^2+\Omega^2}+\frac{g_2\alpha\Omega}{\alpha^2+\Omega^2}}{\frac{g_1\lambda_1^2}{\lambda_1^2+\Omega^
2}+\frac{g_2\alpha^2}{\alpha^2+\Omega^2}}.
\end{equation}
In the above formulas, $\lambda_1$ is given by \cite{hantal90} 
\begin{equation}
\label{lambda1}
\lambda_1 \approx \frac {\sqrt 2}\pi\, (1-\frac 32 D)\,\exp(-1/4D),
\end{equation}
and $\alpha=2$. The weights, $g_1$ and $g_2$ can be obtained from the
expressions
\begin{equation}
\label{weight2}
g_2= \frac {\lambda_1 \langle x^2\rangle_{eq}}{\lambda_1 -\alpha} +
\frac { \langle x^2\rangle_{eq} - \langle x^4\rangle_{eq}}{\lambda_1 -\alpha}
\end{equation}
\begin{equation}
\label{weight1}
g_1=\langle x^2\rangle_{eq} -g_2
\end{equation}
To leading order in $D$, we can replace $\lambda_1$ by $\lambda_K=\sqrt
2/\pi \exp(-1/4D)$, $g_1 \approx 1$ and $g_2 \approx D/ \alpha$. This is
the limit considered in \cite{gang92}.

Linear response theory leads to the following predictions: the first
moment $\langle x(t)\rangle_\infty$ should contain a single harmonics
with the frequency of the driving force, the output amplitude should
behave linearly with $A$.  Certainly, for finite values of $D$, if the
amplitude of the driving force is infinitesimally small, the expansion
procedure in $A$ is valid and LRT applies. The point that we want to
address here is that for finite small amplitudes, $A < A_T$, the value
of $\Omega$ has to be taken into account when applying LRT. The upper
limit for the values of $A$ for which LRT remains valid\footnote{ In the
linear response regime, the dimensionless ratio $A/D$ is assumed to obey
$A < D$. In the opposite singular limit, $A \gg D$, the dynamics assumes
universal weak noise spectral properties \cite{gamhan98,shnjun94}.}
depends as well on the driving frequency.

The adiabatic approximation gives a description of the dynamics when
$\Omega$ is small compared to any other characteristic frequency of the
system. In this approach \cite{junhan91}, the probability density is
assumed to be given by
\begin{equation}
\label{adiab}
P_{ad}(x,t)= N(t) \exp \left (-\frac {U_0(x)-Ax\cos(\Omega t)}{D}\right),
\end{equation}
where $N(t)$ is the normalization constant.  An analysis of the
corrections to the bare adiabatic approximation has been recently
presented by Talkner \cite{talkne99}.

Even in the absence of driving, no exact explicit time dependent
analytical solution of the FPE for the model system at hand is known. We
have resorted to numerical solutions of eq.~(\ref{lfpe}). We
follow a technique based on the use of the split propagator method of
Feit et al. \cite{feifle82} and detailed in \cite{gommor92}. From the
numerical solution of the FPE we can easily obtain the time dependence
of $\langle x(t)\rangle_\infty$. As this is a periodic function of time,
its Fourier components can be obtained by numerical quadrature.

\begin{figure}
\onefigure[scale=0.5]{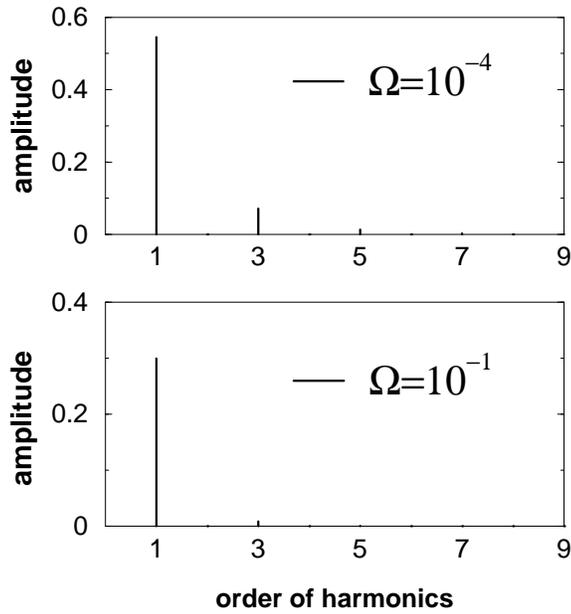}
\caption{Amplitudes of the Fourier components of $\langle
x(t)\rangle_\infty$ for noise strength $D=0.1$ and input amplitude
$A=0.2$ and frequencies  $\Omega=10^{-4}$ (upper panel) and $\Omega=10^{-1}$
(lower panel). }
\label{f.1}
\end{figure} 

In Fig.~\ref{f.1}, we show the amplitudes of the relevant Fourier
components of the output signal for $D=0.1$, $A=0.2$ and two very
different driving frequencies, $\Omega=10^{-1}$ and $\Omega=10^{-4}$.
In this figure, as well as in the subsequent ones, we have taken
$D=0.1$. This is a typical value and it is adequate for the validity of the
two-mode approximation leading to eqs.~(\ref{amp}, \ref{fase}).
On the horizontal axis we indicate the order of
the harmonics. It is clear that even for this driving amplitude,
relatively large in relation to its threshold value, the response of the
system at the larger frequency contains essentially the first
harmonics. On the other hand, for the small driving frequency, higher
order harmonics are generated. This is an indication of the failure of
LRT to describe the dynamics at these low frequencies, while LRT might
still be a good description for higher frequencies.

\begin{figure}
\onefigure[scale=0.5]{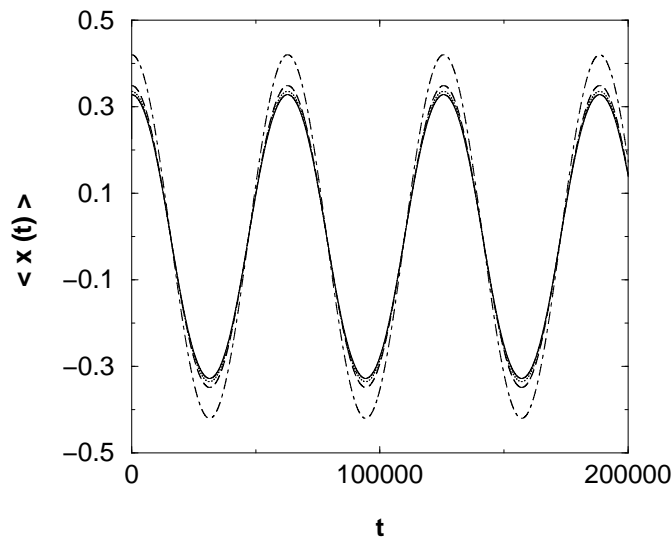}
\caption{Time evolution of $\langle x(t)\rangle$ for $D=0.1$, $A=0.04$
and $\Omega=10^{-4}$ as obtained from the numerical solution of the FPE
(solid line), the adiabatic approximation (dotted line), the two-mode
LRT (dashed line) and the two-mode LRT to leading order in $D$
(dot-dashed line).}
\label{f.2}
\end{figure} 

In Fig.~\ref{f.2}, we depict the time evolution of $\langle x(t)\rangle$
obtained from the numerical solution of the FPE for $D=0.1,\, A=0.04$
and $\Omega=10^{-4}$. We also show the behaviors obtained using the
adiabatic ansatz, eq.~(\ref{adiab}), LRT within the two-mode
approximation, eqs.~(\ref{amp}, \ref{lambda1}, \ref{weight2},
\ref{weight1}) and LRT to leading order in $D$. The input signal is
largely amplified at this small frequency. The adiabatic result deviates
slightly from the numerical one at the peaks. The deviations from the
numerical results are larger with the LRT description. Nonetheless, the
two-mode LRT and the adiabatic approximations yield an acceptable
description of the dynamics. This is expected within the linear response
regime, where $ A \ll D$.

\begin{figure}
\onefigure[scale=0.5]{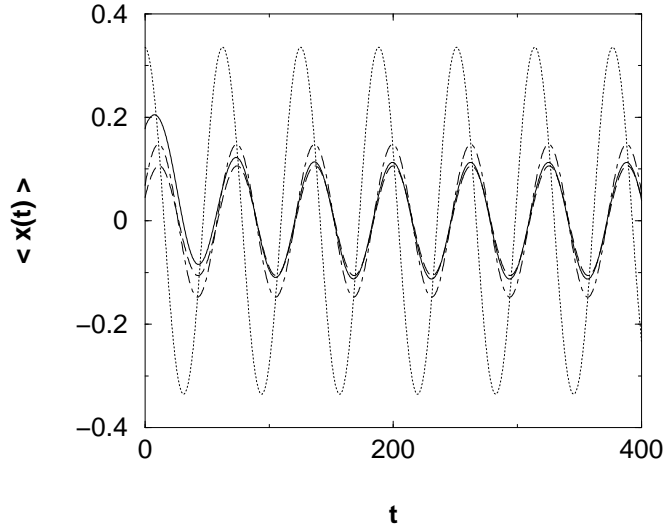}
\caption{The same as in Fig.~\ref{f.2} but with $\Omega=10^{-1}$.}
\label{f.3}
\end{figure}

\begin{figure}
\onefigure[scale=0.5]{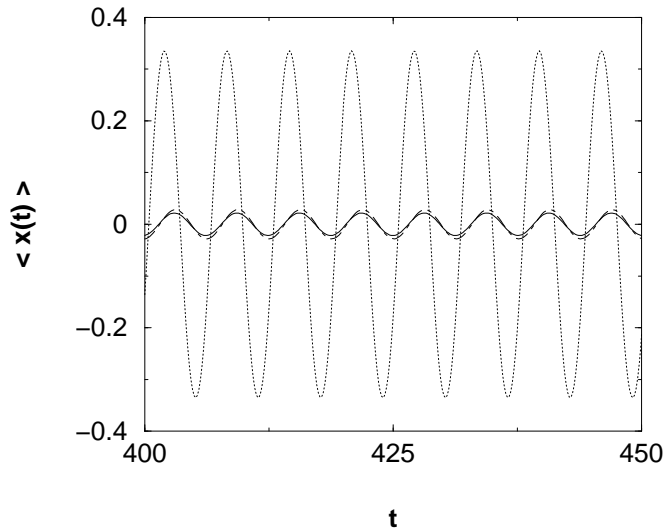}
\caption{The same as in Fig.~\ref{f.2} but with $\Omega=1.0$. Notice
that due to the large value of the driving frequency, we only plot a few
cycles of the output in the asymptotic regime.}
\label{f.4}
\end{figure}

In Fig.~\ref{f.3}, we show the behavior for $\Omega=10^{-1}$. It is
clear that the adiabatic approach yields a signal with a very large
amplitude and a large phase shift compared with the numerics. The
two-mode LRT still yields a very acceptable behavior. The same
qualitative features are observed in Fig.~\ref{f.4} where
$\Omega=1$. For this large frequency, the deviations of the two-mode LRT
from the numerical result are very small and they can not be noticed in
the plot. In Fig.~\ref{f.3}, we show the full time evolution including
the short transient. In Fig.~\ref{f.4}, as we consider a
large frequency value, we only show a few oscillations in the asymptotic
regime, so that the details of a cycle can be distinguished. These last three
figures show that for very small $A$, LRT gives a satisfactory
description of the system response, with deviations from the numerics
more pronounced as the external frequency assumes smaller values.

\begin{figure}
\onefigure[scale=0.5]{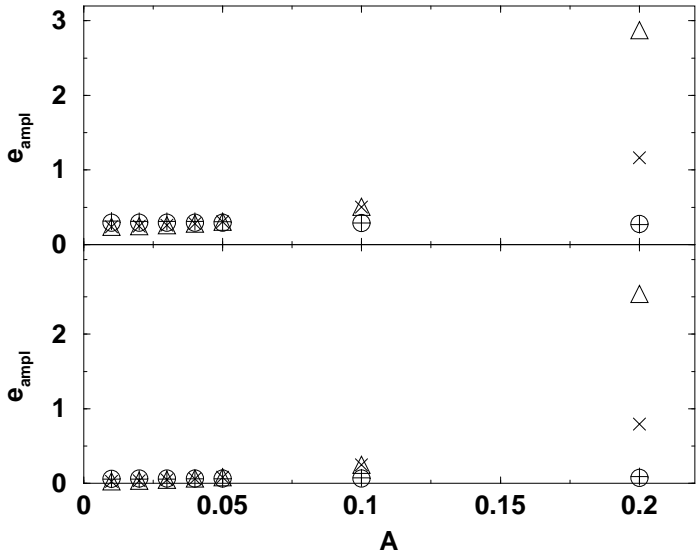}
\caption{Plots of the relative error of the output amplitude, $e_{ampl}
= |A_{out}-a_1|/ A_{out}$, vs. input amplitude $A$ and several values of
the driving frequency. In the upper panel, $a_1$ is evaluated using the
LRT two-mode expressions to leading order in $D$, while the full
two-mode formulas are used in the lower panel; see eqs.~(\ref{amp},
\ref{lambda1}, \ref{weight2}, \ref{weight1}) in the main text. The noise
strength is $D=0.1$ and the frequencies are: $\Omega=1.0$ (circles),
$\Omega=10^{-1}$  (plus signs), $\Omega=10^{-3}$ (x) and
$\Omega=10^{-4}$ (triangles).}
\label{f.5}
\end{figure}

To test the validity of LRT as the input amplitude is increased, we have
carried out an extensive numerical analysis of the system response to
input signals of increasing amplitudes and different frequencies. We
evaluate the relative error $e_{ampl} = |A_{out}-a_1|/ A_{out}$, between
the output amplitude, $A_{out}$, provided by the numerics and the one
obtained within LRT with the two-mode approximation , $a_1$ in
eq.~(\ref{amp}), as a function of the input amplitude $A$. Our findings
are shown in Fig.~\ref{f.5}. The upper panel shows the dependence of
$e_{ampl}$ on $A$ for several frequencies, when the LRT is evaluated to
leading order in $D$, while in the lower panel, the full expressions,
eqs.~(\ref{amp}, \ref{lambda1}, \ref{weight2}, \ref{weight1}) have been
used. For relatively high frequencies, $\Omega=1.0$ (circles), and
$\Omega=10^{-1}$ (plus signs), the error remains small and is
practically constant, even for input amplitudes which are rather large
compared to its threshold value. On the other hand, for small values of
$\Omega$, $\Omega=10^{-3}$ (x) and $\Omega=10^{-4}$ (triangles), the
error increases drastically with the input amplitude. In particular, the
explicit relative errors at $D=0.1$, i.e. ($e_1$, $e_2$, $e_3$, $e_4$),
corresponding to the driving frequencies ($\Omega_1=10^{-4}$,
$\Omega_2=10^{-3}$, $\Omega_3=10^{-1}$, $\Omega_4=1.0$) respectively,
read for $A=0.01$: (0.028, 0.028, 0.056, 0.063); for $A=0.1$: (0.249,
0.249, 0.072, 0.0659); and for $A=0.2$: (2.539, 0.797, 0.090, 0.074).
Thus, the output amplitude predicted by LRT at these small external
frequencies is very much in error, even though, for the same external
amplitudes and moderate-to-large frequencies, LRT predictions are still
adequate.

\begin{figure}
\onefigure[scale=0.5]{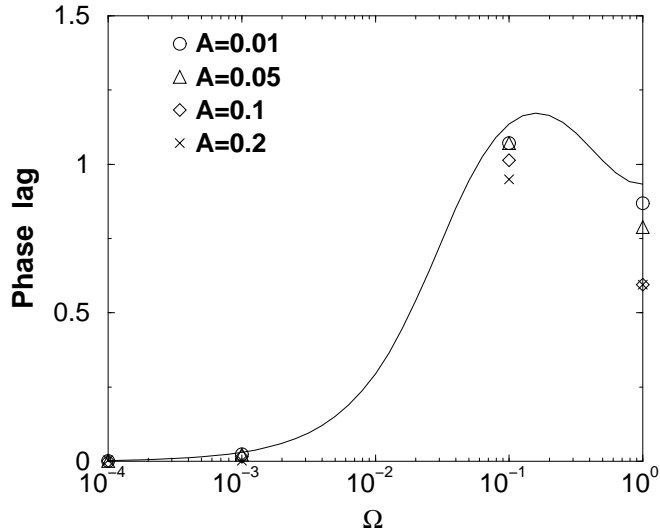}
\caption{Plot of the phase lag between the average output and  the driving
force versus the angular frequency 
$\Omega$. With the solid line we depict $\phi_1^{LRT}$, evaluated using
the LRT two-mode expressions to leading order in $D$ (see
eq.~(\ref{fase}) in the main text).  The symbols denote the numerically
determined values of the phase lag $\Psi$ (see text), for $A=0.01$
(circles), $A=0.05$ (triangles), $A=0.1$ (diamonds), and $A=0.2$ (x).
The noise strength is set at $D=0.1$. }
\label{f.6}
\end{figure}

The average output lags behind the input signal with a phase shift
between $0$ and $\pi/2$. The value of the phase shift predicted by LRT,
$\phi_1^{LRT}$ given by eqs.~(\ref{avg}, \ref{fase}), is independent of
the driving amplitude, but depends on $D$ and $\Omega$. It starts at $0$
for very small frequencies, then reaches a local maximum, and tends to
its limiting value $\pi/2$ for very large frequencies. Its behavior for
the small-to-moderate frequencies considered here ($\Omega < 1 $) is
depicted with the solid line in Fig.~\ref{f.6} for $D=0.1$. On the other
hand, the phase lag of the numerical result, $\Psi$, depends on $D$, $A$
and $\Omega$. The $\Psi$ values plotted have been calculated from the
difference between the instant of times within a period, at which the
driving signal and the periodic output, $\langle x(t) \rangle_\infty$,
cross signs, i.e., the corresponding phase delay in crossing zero. In
Fig.~\ref{f.6}, we plot the values of $\Psi$ for several values of the
driving amplitude and frequency. For very small frequencies, the output
is almost in phase with the input for all the amplitudes considered. As
the frequency increases, deviations between the numerical predictions
and LRT results are manifested, being larger for larger driving
amplitudes.

In conclusion, our analysis clearly indicates the influence of the
driving frequency $\Omega$ on the validity of the LRT predictions for
the amplitude, phase and number of higher harmonics of the response of
the system to subthreshold input signals. As the driving frequency
assumes sufficiently small values, the output amplitude {\it significantly}
deviates from its linear behavior predicted by the two-mode
approximation LRT, even though the driving amplitude might still be
quite small in order to preserve the bistable character of the
unperturbed potential. Even for subthreshold inputs, higher order
harmonics might contribute to the system response for small driving
frequencies, contrary to the predictions of LRT. Although the
global behavior of the phase lag indicated by LRT is qualitatively
correct, as expected, its {\it quantitative} predictions are not reliable as
the input amplitude increases.

\acknowledgments Support by the Direcci\'on General de Ense\~nanza
Superior of Spain (Project No. PB98-1120), the Junta de Andaluc\'\i a
(J.C.-P., J.G.-O., M.M.)  and the Deutsche Forschungsgemeinschaft
HA1517/13-4 (P.H.) is gratefully acknowledged.

\end{document}